\documentclass[journal]{vgtc}                     

\onlineid{1318}

\vgtccategory{Research}

\title{How Good (Or Bad) Are LLMs at Detecting Misleading Visualizations?}

\author{%
  \authororcid{Leo Yu-Ho Lo}{0000-0002-3660-3765} and
  \authororcid{Huamin Qu}{0000-0002-3344-9694}
}

\authorfooter{

  \item
    Leo Yu-Ho Lo and Huamin Qu are with the Hong Kong University of Science and Technology, Hong Kong.
    E-mail: \{yhload@cse, huamin@cse\}.ust.hk.

}

\abstract{%
In this study, we address the growing issue of misleading charts, a prevalent problem that undermines the integrity of information dissemination.
Misleading charts can distort the viewer's perception of data, leading to misinterpretations and decisions based on false information.
The development of effective automatic detection methods for misleading charts is an urgent field of research.
The \leocw{}{recent} advancement of multimodal Large Language Models (LLMs) has introduced a promising direction for addressing this challenge.
We explored the capabilities of these models in analyzing complex charts and assessing the impact of different prompting strategies on the models' analyses.
We utilized a dataset of misleading charts collected from the internet by prior research and crafted nine distinct prompts, ranging from simple to complex, to test the ability of four different multimodal LLMs in detecting over 21 different chart issues.
Through three experiments--from initial exploration to detailed analysis--we progressively gained insights into how to effectively prompt LLMs to identify misleading charts and developed strategies to address the scalability challenges encountered as we expanded our detection range from the initial five issues to 21 issues in the final experiment.
Our findings reveal that multimodal LLMs possess a strong capability for chart comprehension and critical thinking in data interpretation.
There is significant potential in employing multimodal LLMs to counter misleading information by supporting critical thinking and enhancing visualization literacy.
This study demonstrates \leocw{their applicability}{the applicability of LLMs} in addressing the pressing concern of misleading charts.
}


\graphicspath{{figs/}{figures/}{pictures/}{images/}{./}} 

\usepackage{tabu}                      
\usepackage{booktabs}                  
\usepackage{lipsum}                    
\usepackage{mwe}                       

\usepackage{mathptmx}                  
\usepackage{multirow}

\usepackage{rotating}

\newcommand{\eg}{\textit{e.g.}}

\newcommand{\etal}{\textit{et al.}~}
\newcommand{\etalcomma}{\textit{et al.},~}
\newcommand{\fscore}{F1-score}
\newcommand{\fscorespace}{F1-score~}

\newcommand{\osf}{\href{https://osf.io/vx526}{https://osf.io/vx526}}
\newcommand{\squeezeafterfigure}{\vspace{-18pt}}
\newcommand{\squeezeaftertable}{\vspace{-34pt}}
\newcommand{\squeezeaftertablethree}{\vspace{-32pt}}
\newcommand{\squeezebetweenfigureandcaption}{\vspace{-16pt}}
\newcommand{\squeezebetweentableandcaption}{\vspace{-12pt}}
\usepackage[normalem]{ulem} 
\ifx \revision \undefined
    \newcommand{\leo}[2]{{#2}}
    \newcommand{\leocw}[2]{{#2}}
    
    \newcommand{\leonotice}[1]{\empty}
    \newcommand{\leocwnotice}[1]{\empty}
\else
    \newcommand{\leo}[2]{{\sout{#1}\color{orange}{#2}}}
    \ifx \copywrite \undefined
        \newcommand{\leocw}[2]{{#2}}
    \else
        \newcommand{\leocw}[2]{{\sout{#1}\color{blue}{#2}}}
    \fi
    
    \newcommand{\leonotice}[1]{\leo{}{#1}}
    \newcommand{\leocwnotice}[1]{\leocw{}{#1}}
\fi
\ifx \nofigure \undefined
    \newcommand{\getfigure}[1]{#1}
\else
    \newcommand{\getfigure}[1]{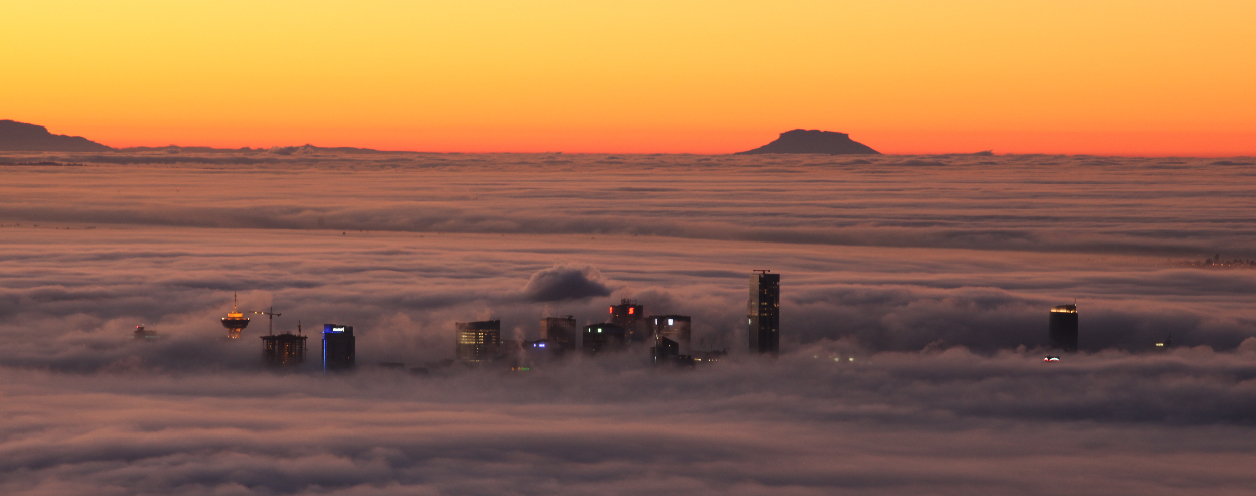}
\fi

\begin{document}


\firstsection{Introduction}

\maketitle


\leonotice{Revision changes are highlighted in orange.}
\leocwnotice{Copywriting changes are highlighted in blue.}

Misleading visualizations have \leocw{consistently}{long} been identified and \leocw{brought into}{included in} public discourse.
In the 1950s, Huff published \textit{How to Lie with Statistics} \cite{huff1993lie}, a book that surfaced the deceptive nature of poorly constructed charts with examples from newspapers.
These charts manipulated the data's visual representation to appear supportive of their intended claims.
Recognizing these discrepancies is crucial when harnessing the power of data visualizations to inform rather than mislead.
Education and critical scrutiny remain the most effective tools for identifying misleading visualizations \cite{chevalier2018observations, bergstrom2017calling}.
Conversely, developing automated tools to detect misleading charts represents a promising area of research.

Recent advancements in the automatic detection of misleading visualizations have largely centered around visualization linters \cite{chen2021vizlinter, lei2023geolinter, mcnutt2018linting}.
These tools scrutinize the structural programming of charts, identifying violations of established visualization guidelines.
Designed to integrate with visualization authoring tools, they alert creators to the potential for misleading content in their charts before publication.
Essentially, these innovations serve as a safeguard on the producer side.
However, the challenge becomes more complex for data consumers, who frequently encounter visualizations in the form of unstructured bitmap images, often embellished with diverse styles and annotations.
This variety poses significant challenges for automated systems, leading to a critical gap: the lack of tools to assist or safeguard the everyday consumers of data visualizations, mainly the general public.
Bridging this gap is crucial, and we urgently need tools to enable data consumers to navigate and interpret data visualizations accurately.

The recent development of Large Language Models (LLMs) \cite{achiam2023gpt} has opened up massive new opportunities, making it possible to tackle complex problems that were once impossible for computer algorithms to solve.
Previous studies have already showcased the remarkable abilities of LLMs to reason logically and interpret data \cite{do2023llms}.
While these models were originally designed for processing and generating text, the introduction of multimodal LLMs has been a pivotal advancement \cite{openai2023gptv} for visualization research.
These multimodal models can understand different types of input, including images, significantly broadening the ways they can be used.
This \leocw{leap forward}{advancement} in \leocw{LLM technology}{LLMs} offers a promising approach to a crucial issue: detecting misleading charts from a consumer's point of view.
The capabilities of multimodal LLMs open up the possibility of creating tools that help people who use data visualizations to better navigate and comprehend visual information, filling a vital need in our digital world.

\textbf{Do multimodal LLMs possess the nuanced understanding required to identify and flag misleading elements in data visualizations?}
To explore this question, our study undertook a thorough assessment of three proprietary \cite{openai2023gptv,ms2023copilot,team2023gemini} and one open-source multimodal LLM \cite{liu2024llavanext}.
Given that the effectiveness of LLMs is significantly shaped by the textual prompts fed \leocw{into }{}them, our initial step was to conduct an exploratory experiment.
This involved crafting three sets of prompts designed to guide the LLMs in recognizing five specific issues within charts.
Pushing the boundaries further, our investigation expanded to assess the LLMs' capabilities in handling an increasing complexity of problems, conducting further experiments that presented the models with charts containing 10 and then 21 different issues\leocw{ in charts}{}.

One of the challenges we faced was the growing difficulty in scaling the number of issues for detection by the LLMs as we broadened the range of issues to be detected.
As we expanded the scope, \leocw{it was inevitable}{we inevitably had} to increase the level of detail of descriptions and instructions, leading to longer prompts and responses.
This escalation resulted in an increase in the length of both the prompts and the generated responses, presenting a significant challenge.
Despite advancements that have expanded the context length LLMs can handle, the processing and generation of lengthy texts remains a computationally intensive and memory-demanding task.
Utilizing the insights gained from our exploratory experiments, our ninth and final prompt was designed to guide the LLMs in detecting 21 different issues.
Instead of designing a single prompt, our final attempt was to generate prompts dynamically, facilitated within a multi-round conversation setup.

Our evaluation revealed the exceptional ability of multimodal LLMs to interpret charts presented as bitmap images.
Following our instructions, these models demonstrated the capabilities of recognizing different chart elements, exercising critical thinking in data interpretation, and detecting a wide range of issues in misleading charts.
Interestingly, LLMs consistently sought additional context for the charts, showcasing an innate caution that proved instrumental in uncovering issues like dubious data sources and concealed information.
Their proficiency in detecting charts with fabricated data was particularly impressive--a challenge that goes beyond structural analysis to require a critical evaluation of the charts' textual content.
These findings showed multimodal LLMs' sophisticated visual understanding and analytical capabilities, unveiling their potential as a powerful resource in creating effective systems for detecting misleading charts.

In a short summary, we aim to provide a glimpse of the potential of \leocw{applying}{utilizing} multimodal LLMs to detect misleading visualizations.
Throughout the study, we share the following results:
\begin{enumerate}
    \item
    Three experiments with a total of nine prompts across different prompting strategies \leocw{covered}{covering} up to 21 charting issues.
    \item
    Evaluation of the prompts on four different multimodal LLMs from proprietary companies and the open-source community.
    \item
    Findings on the challenges of applying LLMs \leocw{on}{to} detecting problematic charts and the strengths and weaknesses of LLMs on similar applications.
\end{enumerate}

All the materials used in the experiments are publicly available on OSF\footnote{\osf\label{osf}}. These include (1) datasets, (2) prompts, (3) program code, (4) conversation logs, and (5) experiment results.

\section{Related Work}

This work intersects four areas of research: (1) Misleading Visualizations, (2) Visualization Linters, (3) Chart Analysis with Computer Vision, and (4) Chart Question Answering with LLMs.

\subsection{Misleading Visualizations}

The discourse on misleading visualizations dates back to the pre-digital era, notably beginning with Darrell Huff's seminal work, \textit{How to Lie with Statistics} \cite{huff1993lie}, in the 1950s.
\leocw{This book}{In this book, Huff} exposed the manipulative potential of data in journalism.
\leocw{This early examination laid the groundwork for Tufte's seminal contributions in the 1980s with his work, \textit{The Visual Display of Quantitative Information}.}{}
\leocw{}{Subsequently, in the book \textit{The Visual Display of Quantitative Information} \cite{tufte1983visual},} Tufte introduced essential concepts like Graphical Integrity and Lie Factors, advancing the \leocw{conversation}{discussion} on how visual data representations can distort the truth.
More recently, Cairo's \textit{How Charts Lie} \cite{cairo2019charts} offers a contemporary viewpoint\leocw{}{ on the issue}, exploring the intricacies of recognizing dubious data sources and the malicious intentions behind visual designs, particularly in the digital media landscape.
These publications provide a comprehensive foundation that informs our understanding of the complexities and ethical issues in data visualization.

Recent academic research has significantly deepened our understanding of misleading visualizations, especially in today's digital era, where misinformation can quickly proliferate online.
Researchers like Pandey \etal \cite{pandey2015deceptive} and Correll \etal \cite{correll2020truncating} have revealed the subtle ways visualizations can be manipulated\leocw{, }{--}for example, by truncating axes\leocw{, }{--}to drastically alter data interpretation.
This expanding body of research emphasizes the urgent need for vigilance in data presentation and perception.
Further studies by Lee \etal \cite{lee2021viral} and Lisnic \etal \cite{lisnic2023misleading} investigate how visualizations contribute to the spread of misinformation on social media and other platforms, underlining the extensive consequences of these deceptive techniques.
In an effort to construct a taxonomy \leocw{on}{of} these deceptions, Lo \etal \cite{lo2022misinformed} compiled a wide range of misleading visualizations from the internet, creating a detailed taxonomy with 74 unique issues.
These academic contributions build upon the discussions initiated by early literature and highlight the continuous challenge and significance of devising methods to counter misinformation in data visualization.
These works offer crucial insights into the mechanics of deceit and the essential role of integrity in data's visual representation.

Data visualization literacy and critical data thinking is an important topic in education.
Chevalier \etal have highlighted the significance of embedding critical thinking within visualization literacy education, advocating for its introduction as early as elementary education \cite{chevalier2018observations}.
This sentiment is echoed and expanded upon at the university level by Bergstrom and West.
In their semester-long course, they instructed students to critically assess the data they encountered, especially through data visualizations.
Their work has culminated in a comprehensive book on the subject \cite{bergstrom2021calling}.
In today's data-driven world, the ability to critically evaluate and interpret information become an essential skill.
Data plays a crucial role in decision-making, influencing policies, and shaping public opinion, highlighting the critical need for individuals to possess the skills to scrutinize data with skepticism and insight.
It's vital to embed critical data thinking within educational curricula and to develop and sharpen tools that enable the public to effectively engage with data visualizations.
Such tools are essential in cultivating an informed society that approaches data visualizations with a critical eye, promoting a more nuanced and discerning consumption of information in our data-intensive reality.


\subsection{Visualization Linters}

In the field of data visualization, significant progress has been made in creating automated systems that are designed to help chart creators produce visualizations that are clear and not misleading.
Inspired by the concept of linters in computer programming—which detect potential errors in code to alert programmers about possible issues leading to incorrect or inefficient outcomes—these visualization linters are a step forward in ensuring the integrity of visual data representations.
McNutt and Kindlmann introduced a linter for matplotlib, a popular charting library, implementing rules derived from the Algebraic Visualization Design (AVD) framework to improve the clarity and effectiveness of charts \cite{mcnutt2018linting}.
Similarly, VizLinter \cite{chen2021vizlinter} uses Answer Set Programming (ASP) to analyze charts created with Vega-Lite, another popular charting library, by scrutinizing chart specifications with established best practices.
Expanding the scope, GeoLinter \cite{lei2023geolinter} applies cartographic principles to map visualizations in Vega-Lite, ensuring that they are not misleading from a cartographic perspective.
These tools \leocw{highlighted}{highlight} the research focus on the production side of data visualization, emphasizing tools that integrate seamlessly with visualization libraries to support chart creators.

While existing research on automated detection systems has largely focused on aiding chart creators in the production process, there is a growing emphasis on empowering consumers to critically evaluate the accuracy and reliability of visual data presentations.
Toward this end, Fan \etal have pioneered a system specifically for analyzing line charts in bitmap format \cite{fan2022annotating}.
This system begins with reverse engineering \cite{poco2017reverse} the chart to extract visualization specifications from its bitmap image.
It then compares these specifications against established best practices, alerting users to any discrepancies or violations directly on the chart image with overlays highlighting misleading aspects alongside corrected versions or annotations, facilitating a more informed interpretation of the data.
Similarly, Hopkins \etal have introduced a visual interface that flags potential issues within charts, directing users' attention to areas that might lead to misinterpretation \cite{hopkins2020visualint}.
Lo \etal analyzed and evaluated six different explanation strategies in explaining common charting issues to the general public \cite{lo2023change}.
These innovations are crucial \leocw{in}{for} improving data literacy among visual data consumers, yet their success heavily relies on the accuracy of chart specification extraction.
Even slight errors at this stage can compromise the system's effectiveness, emphasizing the importance of precision in chart analysis and visualization reverse engineering.

\subsection{Chart Analysis with Computer Vision}

Chart analysis is a fast-developing area within computer vision research focused on extracting data and facilitating question-answering through visual representations.
An initial effort in this field is the creation of the FigureQA dataset, which includes over 100,000 chart images and more than one million binary \leocw{yes or no}{yes-or-no} question-answer pairs, serving as a critical benchmark for evaluating progress in chart analysis \cite{kahou2017figureqa}.
This initiative has encouraged further research, with projects expanding the dataset through synthesis or compilation of real-world charts, enriching the resources for development.
Highlighting the community's dedication to chart analysis, the annual Competition on Harvesting Raw Tables from Infographics \cite{davila2019icdar}, initiated by Davila \etalcomma invites participants to complete seven complex chart analysis tasks.
Together, these tasks form a pipeline that starts \leocw{from}{with} identifying chart types from the chart image, then detecting text, classifying text roles, analyzing axes and legends, identifying plot elements, and finally extracting data.
The competition, with its annotated dataset of over 17,000 synthetic and 16,000 scientific journal-sourced chart images\leocw{.
Such efforts continue to extend}{, continues to push} the limits of computer vision in visualization reverse engineering, improving tools and techniques for machines to understand the vast information in data visualizations.

\leocw{The chart}{Chart} analysis research has rapidly expanded its scope to include sophisticated question-answering setups, moving from simple binary questions to complex open-ended queries that \leocw{required}{require} mathematical reasoning \cite{hoque2022chart}.
This shift marks a substantial advancement in the ability to interpret nuanced data.
During the \leocw{progress}{research}, numerous datasets \leocw{have been}{were} introduced for benchmarking \cite{masry2022chartqa,kafle2018dvqa,methani2020plotqa}.
A common approach to this challenge is treating chart question-answering similarly to table question-answering by extracting the underlying data from charts.
This strategy utilizes advanced techniques developed for table question\leocw{ }{-}answering \cite{masry2022chartqa}.


\subsection{Chart Question Answering with LLMs}

The integration of LLMs has further propelled these advancements, with initiatives like Matcha \cite{liu2022matcha} applying LLMs on code generation to transform chart images back into \leocw{programmatic}{program} code and associated data tables for further processing.
Similarly, UniChart's \cite{masry2023unichart} encoder-decoder framework demonstrates a versatile approach to chart question-answering, potentially bypassing the need for data table generation.
These developments highlight a significant leap toward\leocw{s}{} bridging visual data representations with actionable insights, showcasing the field's progress toward\leocw{s}{} more intuitive and effective data extraction and interpretation methods, thereby enhancing visualization tools for academic research and practical applications.

Effective prompting is the key to successfully integrating LLMs into chart analysis.
An advancement in this area was introduced by Marsh \etal \cite{do2023llms}.
The authors developed a prompt constructor module that effectively bridges the gap between the complex requirements of chart question-answering tasks and the operational capabilities of LLMs.
This module translates these tasks into prompts that enable LLMs to generate relevant and precise outputs. With this strategic prompting mechanism, their system has achieved state-of-the-art results in the field.

Our study builds on this foundation, utilizing multimodal LLMs to identify misleading elements in visual data presentations, a step forward in refining visual data analysis methods and exploring LLMs' potential in mitigating misinformation spread through data visualizations.

\begin{table*}[hbt!]
  \caption[Experiment One Results.]{Experiment One Results:
  Accuracy \leo{}{(Acc.)} represents the ratio of correct answers to the total questions asked.
  The \fscorespace \leo{}{(F1)} is \leo{calculated based on true positives, false positives, and false negatives.}{the harmonic mean of Precision (Prec.) and Recall (Rec.).}
  \leo{}{Across prompts, the models showed higher recall and lower precision, suggesting a tendency toward false positive}
  \leo{}{In Prompt \#2, Copilot chose 3 on a 5-point Likert scale in 78\% of cases, with accurate scores in the remaining six cases (three misleading and three valid).}
  \leo{Relevance indicates if the issue was mentioned in the LLMs' responses.}{Relevance (Rel.) is the percentage of relevant responses.}
  \leo{}{An LLM's response is considered as relevant if the chart issue is mentioned in the response through keyword matching.}
  \leo{In Prompt \#2, Copilot chose 3 on a 5-point Likert scale in 78\% of cases, with accurate scores in the remaining six cases (three misleading and three valid).}{}}
  \squeezebetweentableandcaption
  \centering
  \includegraphics[width=\textwidth, clip, keepaspectratio, trim=0cm 26.3cm 0cm 0cm]{\getfigure{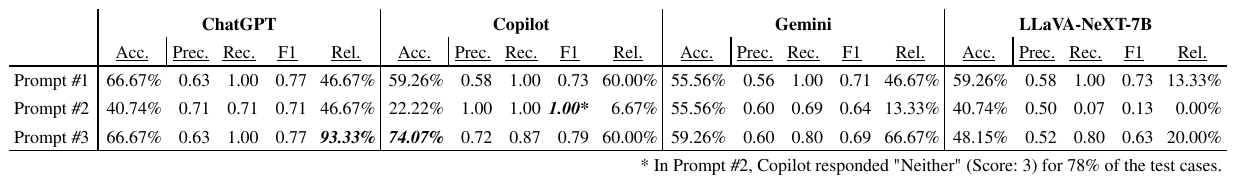}}
  \label{table:exp1}
  \squeezeaftertable
\end{table*}

\section{Methodology}

Our experiments aim to explore the capabilities of multimodal LLMs in identifying misleading visualizations and distinguishing \leocw{these}{them} from more specialized machine learning models.
LLMs are known for their impressive ability to generalize\leo{,}{ and} \leocw{adapting}{adapt} to a wide range of tasks beyond their initial programming through methods like zero-shot, one-shot, or few-shot learning \cite{beltagy2022zero}.
The ability of multimodal LLMs in chart understanding, critically assessing the deceptiveness of these visualizations, and the potential of applying multimodal LLMs in detecting misleading charts are unexplored.
This investigation seeks to determine if such capacities exist and, if so, how crafting prompts alongside the chart images can effectively direct LLMs to produce the intended analytical outcomes.

We began the investigation with an exploratory experiment, collecting various visualizations, including misleading and valid charts.
The objective of the exploration phase is to establish a baseline for evaluating LLMs' ability to distinguish between these two types of charts.
Building upon the insights gained in the exploratory experiment, the second phase of the experiments sharpens the focus on enhancing and broadening the prompts to capture a wider range of chart issues.
However, this expansion poses scalability challenges, particularly in managing the length of the prompts and also the length of the outputs from the LLMs, leading to degraded performance.
To address these issues, the third phase investigates strategies to overcome these limitations, facilitating a more exhaustive examination of the 21 most common issues in charts.
Each stage progressively deepens our understanding of multimodal LLMs' analytical capabilities in detecting visual deceptions.

\textbf{Multimodal LLMs:}
Our \leocw{experimental strategy encompassed a comprehensive evaluation of}{experiments covered} four distinct multimodal LLMs: (1) ChatGPT by OpenAI (gpt-4-1106-vision-preview) \cite{openai2023gptv}, (2) Copilot by Microsoft (model name not specified) \cite{ms2023copilot}, (3) Gemini by Google (gemini-pro-vision) \cite{team2023gemini}, and (4) LLaVA-NeXT (llava-v1.6-mistral-7b-hf) \cite{liu2024llavanext}.
ChatGPT, Copilot, and Gemini are developed by proprietary companies, whereas LLaVA-NeXT is an open-source multimodal LLM.
Notably, at the time of our experiments, LLaVA-NeXT \leocw{is holding}{has achieved} state-of-the-art results across various multimodal LLM benchmarks among open-source LLMs.
Due to hardware constraints, our experiments were conducted with the 7B parameter variant of LLaVA-NeXT, despite the more advanced state-of-the-art 34B parameter model.
Nevertheless, including LLaVA-NeXt in our evaluation covers an important landscape of LLM research.
The LLaVA-NeXt model was run on a server with an Nvidia Titan RTX 24GB graphical processing unit.
On the other hand, the proprietary models were evaluated via their Application Programming Interfaces (APIs). 
We have made the program code of the experiments publicly available on the OSF\footref{osf}.

\textbf{Dataset:}
Our study compiled an evaluation dataset from previous research on misleading charts circulated on the internet \cite{lo2022misinformed}.
The dataset was collected through search engines and social media.
The chart images are annotated with the issues identified from the original web page or social media post.
\leocw{}{A total of }74 unique chart issues were identified in the study, providing a diverse sample of misleading charts that the general public may encounter.

In the initial phase of our exploratory study, we focused on the five most frequently identified issues: (1) Truncated Axis, (2) 3D Chart, (3) Missing Title, (4) Dual Axis, and (5) Misrepresentation.
We subsequently expanded to include the ten most common issues in \leocw{e}{E}xperiment \leocw{t}{T}wo by including (6) Missing Axis Title, (7) Missing Legend, (8) Inconsistent Tick Intervals, (9) Not Data, and (10) Selective Data.
In the third experiment, the study was further broadened to cover up to 21 issues due to a tie for the 20th spot, with both the 20th and 21st issues appearing an equal number of 25 times in the dataset.
These additional issues included (11) Dubious Data, (12) Missing Value Labels, (13) Area Encoding, (14) Overusing Colors, (15) Inappropriate Axis Range, (16) Indistinguishable Colors, (17) Ineffective Color Scheme, (18) Discretized Continuous Variable, (19) Missing Normalization, (20) Missing Axis, and (21) Inconsistent Binning Size.
For each issue, we \leocw{selected}{randomly sampled} six images, dividing them equally into development and test sets, resulting in sets for evaluation across five (N=30), ten (N=60), and 21 issues (N=126).

Moreover, we included a set of valid charts to evaluate if LLMs can distinguish between misleading and accurate representations. 
This valid set, \leocw{}{randomly }sampled from a collection by Brokin \etal \cite{borkin2013makes}, consisted of chart images from various sources, including news and governmental organizations.
Through stratified sampling from this pool of 2,000 chart images--excluding infographics and charts with issues like missing titles or 3D effects--we assembled a valid chart set of 24 images.
This set was evenly sampled from four sources, each contributing six images, and covered six different chart types, each represented by four images.
This sampling method ensured wide coverage of sources and chart types, minimizing bias toward\leocw{s}{} particular charting styles, data types, or chart types.
Unlike the misleading chart set, which was expanded throughout the progression of our experiments, the valid chart set remained the same throughout the study.
The valid set is divided evenly into development and test sets while maintaining its stratified property.

For each of our experiments, we organized the datasets as follows: (1) for Experiment One, we had a total of 54 images ($N=54$), with 27 in the development set ($N_{dev}=27$) and 27 in the test set ($N_{test}=27$); (2) for Experiment Two, the total was 84 images ($N=84$), split equally into 42 for development ($N_{dev}=42$) and 42 for testing ($N_{test}=42$); and (3) for Experiment Three, we worked with 150 images ($N=150$), divided into 75 for development ($N_{dev}=75$) and 75 for testing ($N_{test}=75$).
Unlike traditional machine learning models, our study lacked a training phase, hence the development set was used for creating prompts, and the test set was strictly for evaluation purposes.

\textbf{Evaluation Metrics:}
Throughout the various stages of our experiments, as we gained additional insights into the behavior of the LLMs, we fine-tuned our evaluation metrics accordingly.
In the initial exploratory study, we employed accuracy and \fscorespace as indicators of classification performance, alongside keyword matching on the reported issues, to assess the relevance of the generated outputs.
From \leocw{p}{P}rompt \#3 of Experiment One and after, we discovered that the LLMs consistently followed our naming of each issue.
Therefore, exact matching is used in these experiments, and keyword matching is only used for Prompt \#1 and Prompt \#2 of Experiment One.
\leo{}{If the chart issue is mentioned in the response, we consider the response as relevant, and vice versa.}

In order to process the LLM outputs, after each prompt, we \leocw{feed}{fed} the prompt and output as a conversation dialogue into the LLM and ask it to summarize the dialogue in JSON format.
The reported results are based on this JSON instead of the free-formed responses.

As we progressed to Experiment Two and Experiment Three, expanding the misleading charts set led to an imbalance between the number of misleading and valid charts.
We recognized that accuracy was inadequate as a measure of performance.
Consequently, we rely on the \fscorespace in Experiment Two and Experiment Three.
Additionally, beginning with the third prompt, we introduced factual questions about the charts into our queries to the LLMs.
This enabled us to directly measure the LLMs' capabilities in chart comprehension.
Therefore, we \leo{}{labeled these chart properties manually on all the 150 chart images and }included factual correctness as a metric for evaluation in Experiment Two and Experiment Three.

Overall, we \leocw{have }{}conducted three experiments, each with a progressively larger set of issues to cover and posed different sets of challenges.
We employed different prompting strategies in different stages to effectively instruct the LLMs to detect misleading visualizations.

\section{Experiment One: Exploration Phase}

\begin{sidewaysfigure*}
  \centering
  \includegraphics[width=\linewidth, height=\textheight, keepaspectratio, clip, trim=0.19cm 0.9cm 0.19cm 0.19cm]{\getfigure{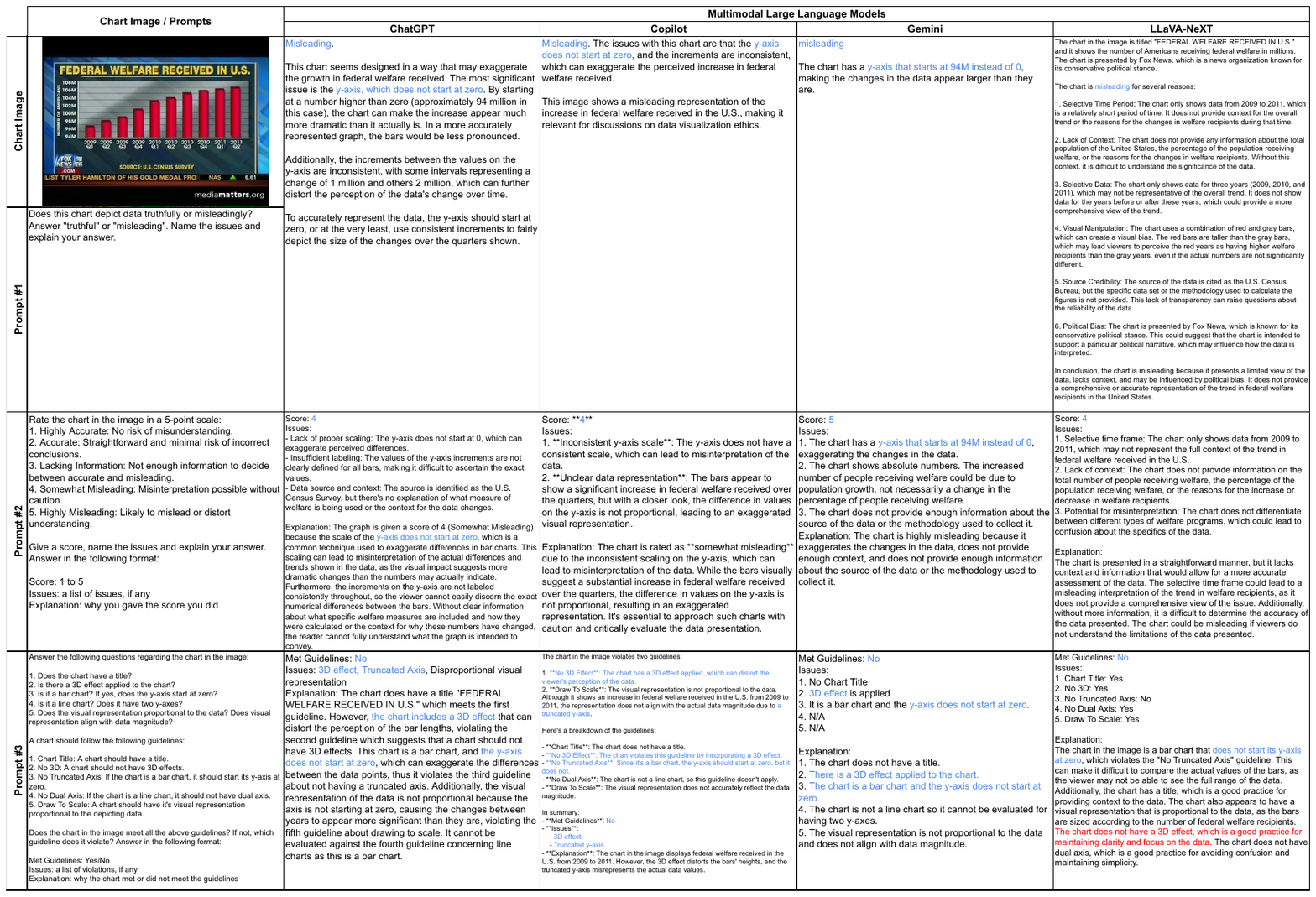}}
  \squeezebetweenfigureandcaption
  \caption[Experiment One Prompts.]{Experiment One Prompts:
  Prompt \#1 directly asks if the chart is misleading or truthful.
  Prompt \#2 requests a rating on a 5-point Likert scale, with 3 being neutral.
  Prompt \#3 applies the Chain of Thought\leocw{s}{} strategy for a structured reasoning process with LLMs.
  The example chart has a major issue of truncated axis and a minor issue of 3D effects on the bars.
  Phrases in \textcolor[HTML]{487CDF}{blue denote accurate interpretations}, and phrases in \textcolor{red}{red indicate incorrect interpretations}.}
  \label{fig:exp1}
\end{sidewaysfigure*}

Our understanding of how to prompt multimodal LLMs to identify misleading charts—a task they have not previously encountered—is limited.
Thus, we \leocw{begin}{began} our exploration with a straightforward prompt: ``Does this chart depict data truthfully or misleadingly?''.
Through this approach, we aim to incrementally gain insights into effectively interacting with LLMs for the purpose of developing an automated system capable of detecting misleading visualizations.

\begin{table*}[hbt!]
  \caption[Experiment Two Results.]{Experiment Two Results:
  The \fscorespace \leo{ assesses}{(F1), precision (Prec.), and recall (Rec.) assess the LLMs'} performance in an imbalanced setting ($N_{misleading}=30,N_{valid}=12$).
  \leo{}{Compared to Experiment One, the high-recall, low-precision situation has been improved but we are still observing large percentage of false positives.}
  \leo{Relevance indicates if the issue was mentioned in the LLMs' responses.}{Relevance (Rel.) is the percentage of relevant responses.}
  \leo{}{An LLM's response is considered as relevant if the chart issue is reported as one of the pitfalls.}
  \leo{Factual Correctness measures the percentage of correct answers to the factual questions about chart type, axis, scale, and encoding.}{Factual Correctness (Fact.) measures the percentage of correct answers to the factual questions about the chart, \eg, chart type, title, axis, scale, and encoding.}}
  \squeezebetweentableandcaption
  \centering
  \includegraphics[width=\textwidth, clip, keepaspectratio, trim=0cm 26.75cm 0cm 0cm]{\getfigure{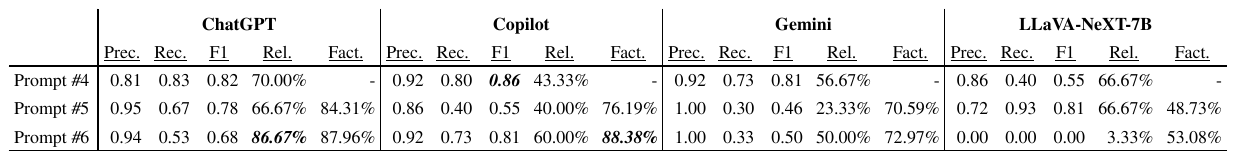}}
  \label{table:exp2}
  \squeezeaftertable
\end{table*}


To initiate this exploration, we \leocw{start}{started with} a small set of misleading ($N_{misleading}=30$) and valid ($N_{valid}=24$) charts divided equally into a development set and a test set.
To analyze the results objectively, without relying on our interpretation of the text generated by the LLMs, we re-input the dialogues generated in response to each prompt back into the LLMs, requesting \leocw{that they}{the LLMs to} summarize the dialogue in JSON format.
This method allows for a structured and unbiased analysis of the LLMs' responses.

In this preliminary experiment, we tested three types of prompts: (1) direct inquiry, (2) Likert scale assessment, and (3) Chain of Thought (CoT) reasoning \cite{wei2022chain}.
We evaluated the outcomes based on accuracy, \fscore, and relevance of the responses generated by the models.

\textbf{Prompt \#1 (Direct Ask):}
The initial prompt we experimented with was directly asking: ``Does this chart depict data truthfully or misleadingly?''
When applied through web user interfaces, the responses often elaborated on the characteristics of misleading visualizations and provided a checklist rather than offering a direct ``truthful'' or ``misleading'' answer.
To guide the responses more effectively, we appended the instruction\leocw{}{ with} ``Answer `truthful' or `misleading'. Name the issues and explain your answer.'' following the initial question.
After receiving each response, we formatted the question and response into a dialogue and requested that the LLMs summarize this dialogue in JSON format.
An illustration of various prompts and responses from different LLMs is presented in \cref{fig:exp1}.

\textbf{Results of Prompt \#1:}
In the test set comprising 27 responses, most LLMs labeled the charts as ``misleading,'' \leo{with}{resulting in high recalls and low precisions.}
ChatGPT \leocw{showing}{showed} a marginally better capability by correctly identifying three out of 12 valid charts.
The issue of misrepresentation was the most commonly cited problem, with responses noting concerns such as ``lack of scale consistency'' and ``disproportional representation.''
\cref{table:exp1} shows the results of Experiment One.

Despite the tendency of LLMs to criticize charts excessively, the results were encouraging in terms of their ability to identify certain issues, such as the inappropriate use of 3D effects, dual axes, and truncated axes.
On the other hand, the absence of titles on charts did not seem to be flagged as a problem by any of the LLM responses.
While not directly misleading, the omission of a title results in a loss of context, rendering the chart less comprehensible.
The concept of ``lacking context'' was, however, frequently mentioned in the feedback.
\cref{fig:issues} shows the summary of the issues recognized by the LLMs across different prompts.

\textbf{Prompt \#2 (Likert Scale):}
Based on the findings \leocw{in}{of} \leocw{p}{P}rompt \#1, we considered introducing more nuance to the responses using a 5-point Likert scale, where the midpoint represents insufficient information to make a definitive judgment.
This adjustment allows the LLMs to provide answers that aren't strictly ``truthful'' or ``misleading.''
We grouped their responses into three categories: scores of \leocw{1-2}{1 or 2} as ``truthful,'' a score of 3 as ``neither,'' and scores of \leocw{4-5}{4 or 5} as ``misleading.''
In addition to accuracy and the \fscore, it became necessary to track the proportion of responses categorized as ``neither.''

\textbf{Results of Prompt \#2:}
When utilizing this more nuanced prompt, all LLMs exhibited a decline in performance compared to the initial prompt, with the exception of Copilot, which achieved a perfect \fscore.
Copilot's responses were particularly notable, with 21 out of 27 categorized as ``neither,'' alongside three correctly identified as misleading and three as truthful.
Despite this, the relevance of the responses to the specific issues present in the charts was minimal.
Overall, using a Likert scale prompt did not effectively guide LLMs in pinpointing issues within charts.

\textbf{Prompt \#3 (Chain of Thought):}
In the third prompt, we implemented the CoT prompting strategy, which encourages LLMs to undergo a reasoning process before delivering a definitive answer.
This approach is inspired by the methodology that guides LLMs through a step-by-step analytical process, culminating in a reasoned conclusion \cite{wei2022chain}.

The prompt was designed to inquire about critical aspects of potential charting issues, such as the presence of a title, the use of 3D effects, and whether the axis begins at zero, among others.
Subsequently, we presented a set of guidelines and questioned whether the chart adhered to all these criteria.

\textbf{Results of Prompt \#3:}
Despite the \fscore for this prompt being similar to those of the first prompt, there was a notable enhancement in the relevance of the responses.
By providing clear instructions, the LLMs were able to demonstrate an understanding of the charts and accurately identify issues as directed.
This indicates that, with precise guidance, LLMs can effectively engage in the task of chart analysis.

\textbf{Discussion of Experiment One:}
In this exploratory experiment, we established an experimental setup to assess LLMs' abilities in chart comprehension, critical thinking, and adherence to instructions.
Traditional algorithmic detection systems often yield straightforward \leocw{yes or no}{yes-or-no} answers.
However, the LLM responses were notably non-binary.
For example, in response to \leocw{p}{P}rompt \#2, ``it depends'' was a frequent reply.
This tendency aligns with advancements in LLM training, particularly the emphasis on generating human-like responses through reinforcement learning based on human feedback \cite{ouyang2022training}.
A plausible reason for this pattern is the preference for ``it depends'' responses by human annotators over unequivocal but potentially incorrect answers.

We introduced a checklist with issue names to reduce the ambiguity in parsing responses.
For \leocw{p}{P}rompts \#1 and \#2, keyword matching was employed to determine if the responses addressed the identified issues within the charts.
By contrast, in \leocw{p}{P}rompt \#3, we provided a checklist, and the LLMs consistently used the specified terminology in their responses.
This approach significantly clarified the evaluation of \leocw{}{the }responses' relevance, making the results from \leocw{p}{P}rompt \#3 much clearer than those from the earlier prompts.

For our subsequent experiment, rather than seeking a simple \leocw{yes or no}{yes-or-no} verdict, we \leocw{will instruct}{instructed} the LLMs to categorize identified issues by their severity, including an option for inconclusive findings.

\section{Experiment Two: Consolidation and Expansion Phase}

\begin{figure*}[hbt!]
  \centering
  \includegraphics[width=\linewidth, height=\textheight, keepaspectratio, clip, trim=0.19cm 14.8cm 0.19cm 0.19cm]{\getfigure{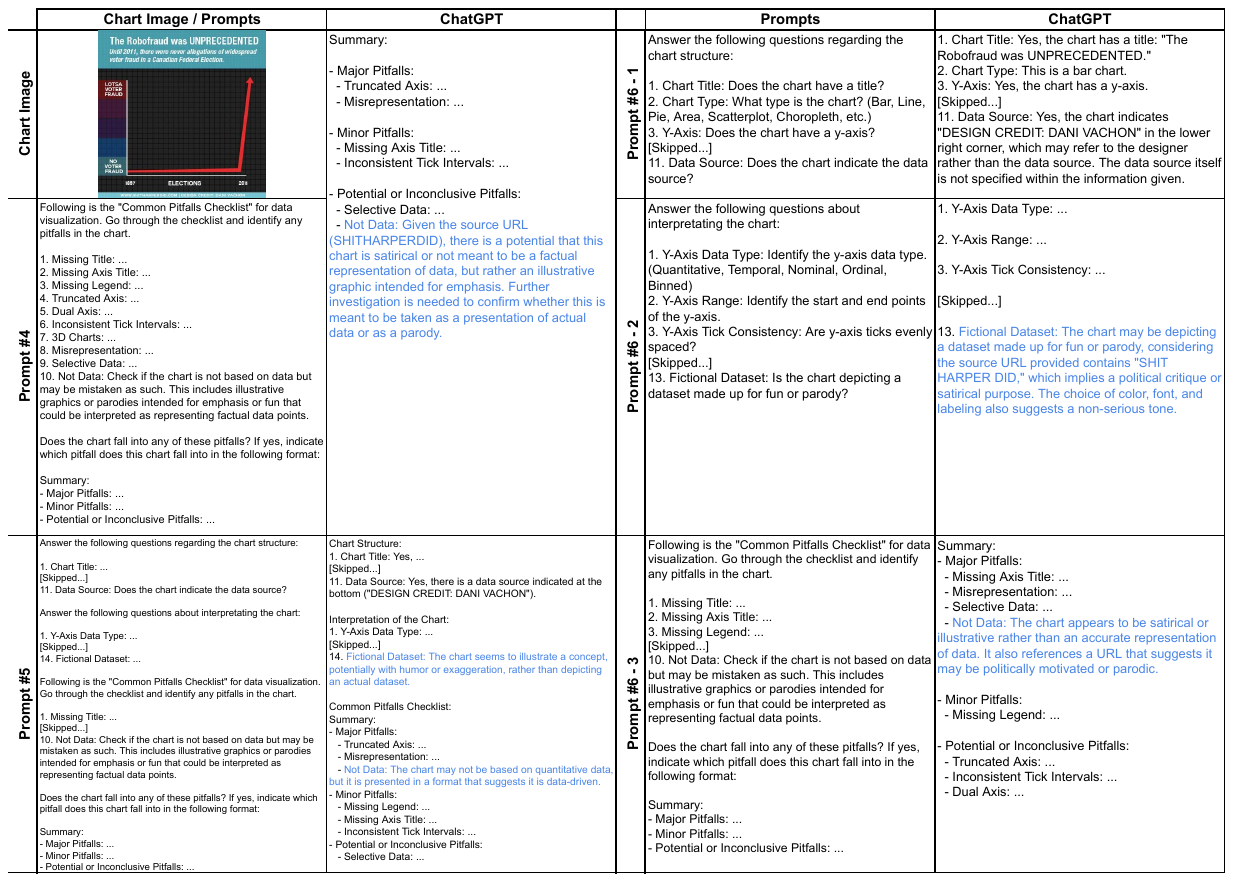}}
  \squeezebetweenfigureandcaption
  \caption[Experiment Two Prompts.]{Experiment Two Prompts:
  Prompt \#4 poses a checklist for LLMs to go through.
  Prompt \#5 applies the Chain of Thought\leocw{s}{} strategy.
  Prompt \#6 modifies Prompt \#5 to prevent LLMs from bypassing reasoning questions.
  The example chart contains fictional data for parody purposes.
  Phrases in \textcolor[HTML]{487CDF}{blue denote accurate interpretations}. Portions of the prompt and responses are omitted for clarity.}
  \label{fig:exp2}
  \squeezeafterfigure
\end{figure*}

In the first experiment, we derived three key insights regarding the efficacy of prompting LLMs to perform the task of detecting misleading charts: (1) posing factual questions and \leocw{}{providing }a checklist \leocw{is}{are} beneficial, (2) LLMs tend to avoid giving definitive answers, and (3) LLMs followed our naming of the issues.

In \leocw{e}{E}xperiment \leocw{o}{O}ne, we found that the third prompt, which utilized the CoT strategy, was particularly effective in guiding LLMs to pinpoint issues in charts.
While the CoT strategy has proven effective in previous studies \cite{wei2022chain}, our construction of \leocw{p}{P}rompt \#3 included reasoning steps and a checklist of potential chart issues.
We are interested in distinguishing between the impacts of the checklist and the CoT strategy.
To explore this, \leocw{p}{P}rompt \#5 replicates the approach of \leocw{p}{P}rompt \#3, and \leocw{p}{P}rompt \#4 employs only a checklist, omitting the CoT reasoning steps.

As the experiment progressed to include more issues, the \leocw{prompts' length}{length of prompts}, incorporating extensive issue definitions and related factual questions, hinted at potential scalability challenges.
Particularly, \leocw{p}{P}rompt \#5 revealed that posing numerous questions within a single prompt might lead LLMs to overlook earlier questions, focusing instead on the final checklist question only.
To address this, \leocw{p}{P}rompt \#6 adopts a multi-pass conversation approach, dividing the extensive \leocw{p}{P}rompt \#5 into smaller prompts within a continuous dialogue format.
This forces the LLMs to answer all the factual questions and reasoning steps before jumping to the checklist question to give the final answer.

In the second experiment, we expanded the scope to include the ten most common issues, thereby enlarging the dataset of misleading charts.
The dataset for this phase comprised 60 misleading charts with the same set of 24 valid charts in the first experiment.
Both sets are evenly divided into development and testing sets. 

Given the dataset's imbalance between misleading and valid charts, we determined that accuracy was inadequate and opted to rely on the \fscorespace instead.
In \leocw{p}{P}rompts \#5 and \#6, the LLMs were prompted to include the final judgment of chart issues and the answers to the factual questions.
Therefore, for \leocw{p}{P}rompts \#5 and \#6, we also evaluated the accuracy of responses to factual questions.
\cref{fig:exp2} shows the prompts evaluated in Experiment Two.

\textbf{Prompt \#4 (Checklist Only):}
We instructed LLMs to go through a checklist and categorize identified issues into three levels of concern: major, minor, and either potential or inconclusive pitfalls.

\textbf{Prompt \#5 (Chain of Thought):}
The prompt was structured on top of the checklist in Prompt \#4 in a tiered manner, divided into three segments.
The first segment posed basic inquiries regarding the chart's structure, including the type of chart, presence of 3D effects, axis, legends, and data source.
The second tier involved more complex questions necessitating interpretation, such as determining the data type represented on the axis, the range of the axis, and whether the axis ticks were evenly spaced.
Given the observed ability of LLMs to contextualize their analysis based on chart content, we \leocw{also inquired}{investigated} whether the information \leocw{provided}{presented} in the chart \leocw{}{alone }was sufficient for understanding\leocw{}{,} or if additional details\leocw{, like a}{ such as a chart} title or axis titles\leocw{,}{} were necessary.
The final set of questions, addressing the data directly, was the most intricate, probing whether the data might have been selectively chosen or fabricated.

\textbf{Prompt \#6 (Split Chain of Thought):}
During the execution of \leocw{p}{P}rompt \#5, we noted instances where the LLMs' generated responses either reached the output limit or the models bypassed the initial two sets of questions, opting to respond directly to the final checklist.
To address this, we segmented the prompt into three parts, presenting them to the LLMs in successive rounds while incorporating the preceding dialogue.
This method allowed for a more manageable and structured approach to eliciting detailed analyses from the LLMs.

\begin{figure}[hbt!]
  \centering
  \includegraphics[width=\linewidth, height=\textheight, keepaspectratio, clip, trim=0.19cm 7.2cm 0.19cm 0.19cm]{\getfigure{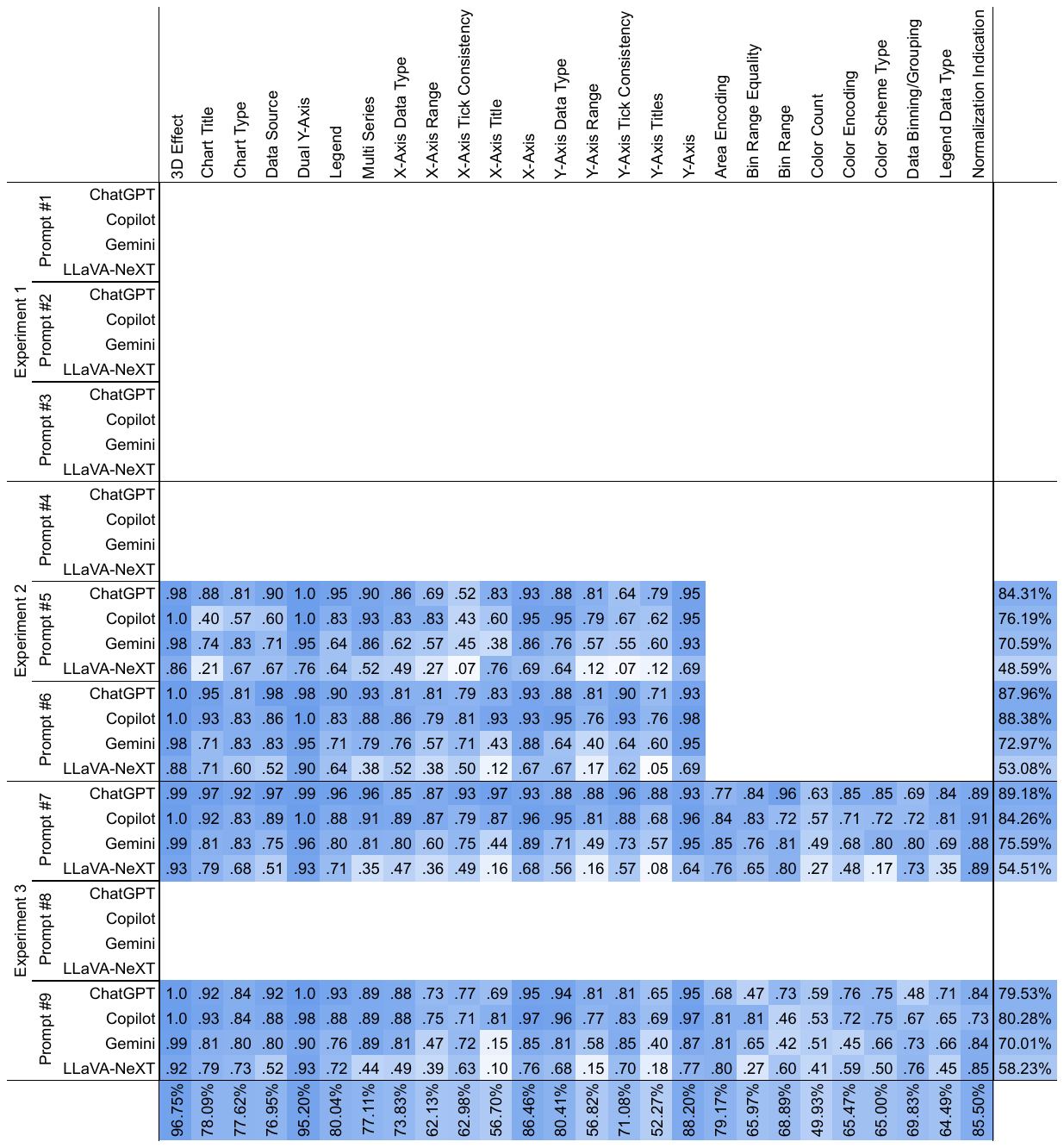}}
  \squeezebetweenfigureandcaption
  \caption[Percentage of factual questions accurately answered by LLMs.]{Percentage of factual questions accurately answered by LLMs on chart properties, axes, scale, and encoding.
  Edge numbers indicate the correct answer percentage for each row or column.}
  \label{fig:factual}
  \squeezeafterfigure
\end{figure}

\textbf{Experiment Two Results and Discussion}
In categorizing the responses, we classified charts as ``misleading'' if they presented at least one major issue; otherwise, they were deemed valid.
The outcomes of \leocw{e}{E}xperiment \leocw{t}{T}wo are summarized in \cref{table:exp2}.
While the three prompts yielded \leocw{}{a }similar \fscore, the quantity and relevance of the issues identified varied across different prompts.
Specifically, LLMs reported a greater number of irrelevant issues when responding to the checklist in \leocw{p}{P}rompt \#4.
In contrast, when engaging with the factual questions in \leocw{p}{P}rompts \#5 and \#6, the incidence of reported issues decreased, yet the mention of relevant issues remained fairly consistent.
Regarding these factual questions, \leocw{}{the }LLMs demonstrated a surprising \leocw{capability of understanding}{ability to understand} the charts, accurately answering questions about recognizing different chart components, and correctly interpreting scales and encodings.
This was achieved without specific training or one-shot or few-shot demonstrations, relying solely on the instructions within the prompts, showcasing the LLMs' impressive capacity for visual interpretation.
\cref{fig:factual} shows the results of \leocw{}{the }factual questions answered by \leocw{}{the }LLMs.

An interesting discovery from this experiment was the LLMs' ability to identify charts intended as parodies created with fictitious data rather than genuine information.
\cref{fig:exp2} shows an example \leocw{charts}{chart} that \leocw{were}{was} flagged as ``Not Data,'' with advisories against considering \leocw{them}{it} as factual.

During this experiment, it was noted that LLaVA-NeXT experienced issues with repetitive outputs, echoing certain sentences until they reached the maximum token count and terminated, leading to outputs that contained considerable repetitions.
This significantly impacted its performance.
Similarly, Gemini indicated errors suggesting \leocw{it, too}{that it} faced issues with repetitions\leocw{}{, too}.
To address these erratic responses, we adjusted the temperature setting of Gemini, which, as per API documentation, influences the randomness of the response.
In other cases, generating long output\leocw{}{s} triggered the maximum token limitation and prematurely returned incomplete responses.
We simply \leocw{rerun}{reran} the experiment with the same prompt and image for these cases.


Experiment \leocw{t}{T}wo highlighted a limitation of LLMs: as the number of issues to be detected increases, so does the length of the prompts.
For LLMs with input length restrictions, this poses a scalability challenge.
Additionally, more comprehensive prompts lead to longer responses, potentially exceeding the maximum response length the model can generate.
For instance, the ChatGPT API has a limit of 5000 tokens, while Gemini's limit is 2048 tokens on the newly generated text.
The complexity of prompts also increases the likelihood of repetitive output issues.

Addressing the prompt length constraint, we found that while the checklist in \leocw{p}{P}rompt \#4 was shorter, it also prompted LLMs to report more irrelevant issues.
Conversely, \leocw{p}{P}rompt \#6 demonstrated a successful workaround, yielding relevant results and improved factual accuracy.

\section{Experiment Three: Tackling Scalability Issues}

\begin{figure*}[hbt!]
  \centering
  \includegraphics[width=\linewidth, height=\textheight, keepaspectratio, clip, trim=0.19cm 14.8cm 0.19cm 0.19cm]{\getfigure{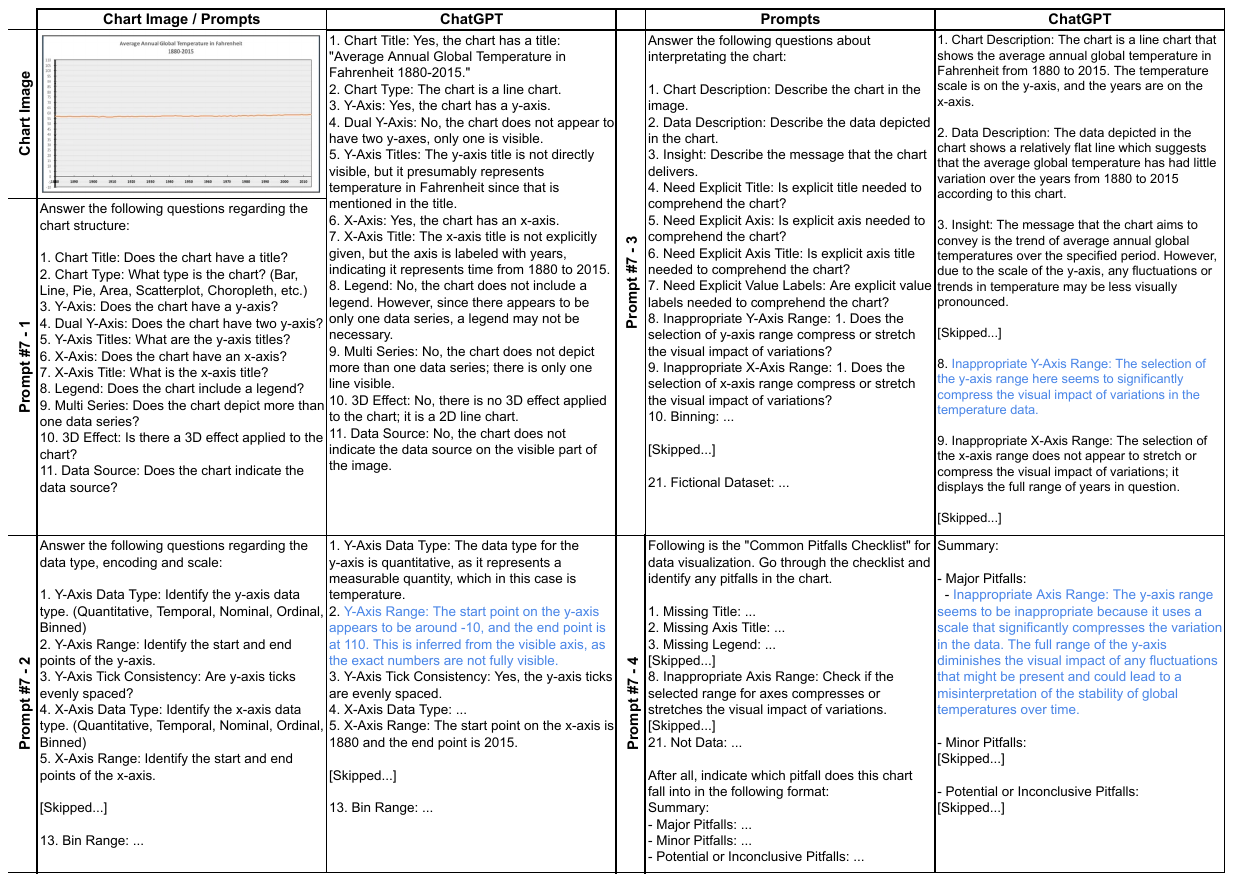}}
  \squeezebetweenfigureandcaption
  \caption[Experiment Three Prompts.]{Experiment Three Prompt \#7 extends Prompt \#6 in applying the Chain of Thought strategy to include additional chart issue definitions.
  The example chart has a major issue of setting the y-axis range inappropriately.
  Phrases in \textcolor[HTML]{487CDF}{blue denote accurate interpretations}. Portions of the prompt and responses are omitted for clarity.}
  \label{fig:exp3}
  \squeezeafterfigure
\end{figure*}

In Experiment Two, expanding the scope to cover ten issues revealed challenges associated with lengthy prompts or responses, leading to premature termination and omission of reasoning questions.
Additionally, despite LLMs correctly answering the factual questions in Experiment Two, they occasionally made contradictory decisions in their final assessments, such as noting a missing title but not recognizing it as an issue in their conclusive response.
We incorporated more interpretive questions within the prompts to refine this aspect in Experiment Three.

\textbf{Prompt \#7 (Split Chain of Thought):}
The initial layer consisted of basic factual inquiries, followed by a second layer examining data types, encoding, and scales, akin to the approach in \leocw{p}{P}rompt \#6.
The third layer required LLMs to evaluate the chart more critically, considering the necessity of explicit titles and axes or the suitability of the axis range and color scheme.
The strategy of segmenting prompts into multiple conversational passes proved effective in Experiment Two.
Yet, as we sought to identify a broader range of 21 chart issues in Experiment Three, the length of each segmented prompt also increased considerably.
The resulting full prompt comprises approximately 6,500 characters and totaling 1,550 tokens.
\cref{fig:exp3} shows an example of Prompt \#7.


\textbf{Prompt \#8 (Direct JSON Output):}
Directly instructing LLMs to go through a checklist was deemed ineffective based on findings from Experiment Two.
In Experiment Three, we tried to condense responses while maintaining the reasoning steps by instructing LLMs to produce a direct issue list in JSON format.

\textbf{Prompt \#9 (Dynamic Chain of Thought):}
Another approach involved dynamically constructing split prompts based on LLM responses, starting with basic questions about chart properties and tailoring follow-up queries based on these initial answers.
For instance, if an LLM identified a chart as a pie chart, subsequent questions about axis range or tick consistency were deemed irrelevant and omitted from the follow-up prompt.



\begin{figure}[hbt!]
  \centering
  \includegraphics[width=\linewidth, height=\textheight, keepaspectratio, clip, trim=0.19cm 3cm 2.7cm 0.19cm]{\getfigure{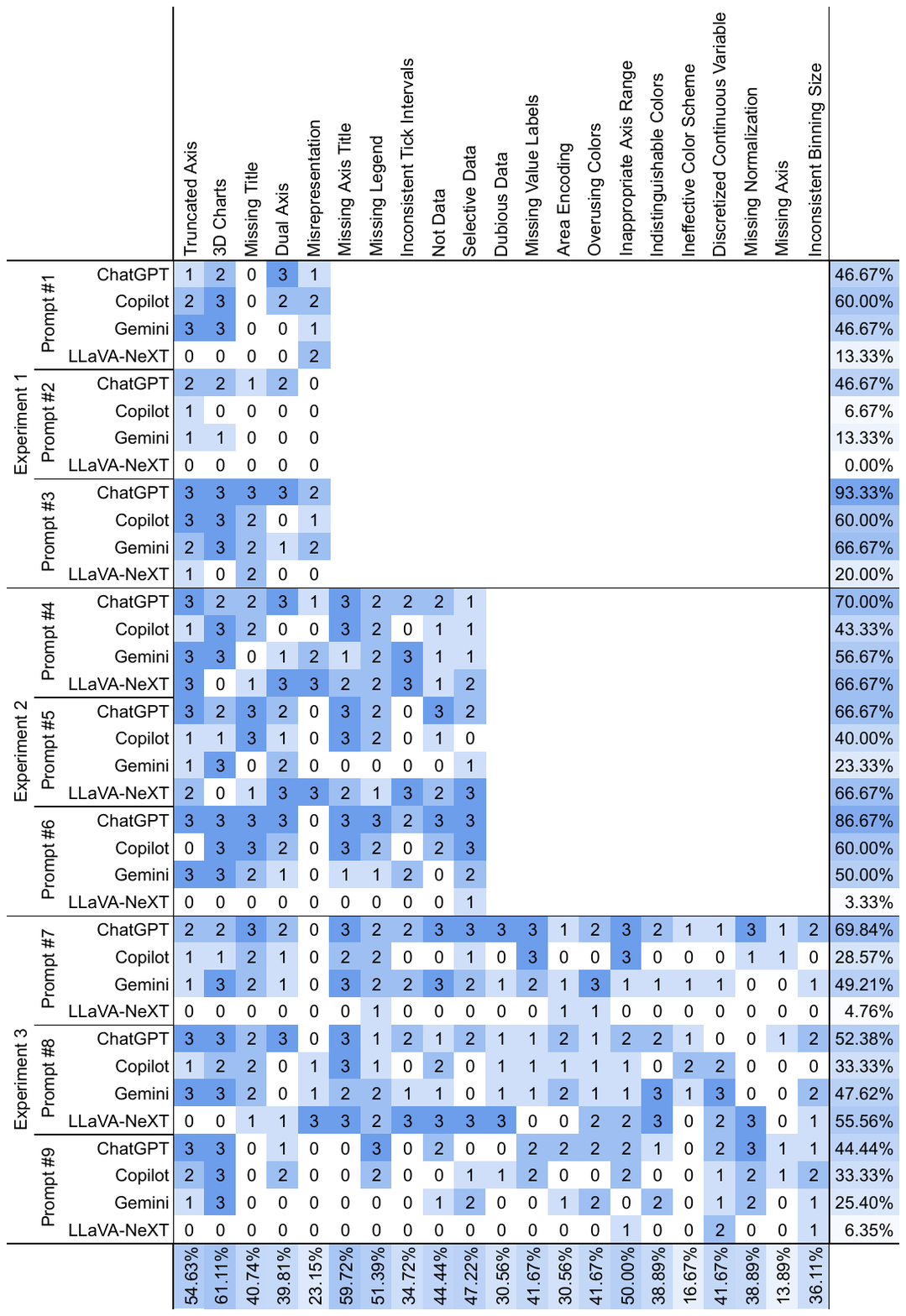}}
  \squeezebetweenfigureandcaption
  \caption[Issues detected by LLMs across different prompts.]{Issues detected by LLMs across different prompts, each issue appeared three times in the test set.
  Edge numbers indicate the correct identification percentage for each row or column.}
  \label{fig:issues}
  \squeezeafterfigure
\end{figure}

\textbf{Experiment Three Results and Discussion:}
Prompt \#8 encountered issues similar to those observed with \leocw{p}{P}rompt \#4, with a tendency to over-report issues in the charts.
Specifically, when using \leocw{p}{P}rompt \#8, despite a high \fscore, Copilot identified major issues in 74 charts, deeming only one chart as valid, which was inaccurately classified as a false negative.
The \fscore, generally appropriate for datasets with an imbalance between categories, was utilized for evaluation.
The test dataset for this phase included 63 misleading charts and 12 valid charts.
Other than prompt \#8, both prompts \#7 and \#9 showed more balanced performance, correctly identifying valid charts (true negatives) and accurately highlighting issues in misleading charts (true positives).
\cref{table:exp3} shows the results of Experiment Three.

Prompt \#9, designed to reduce the number of excessively reported issues by filtering the checklist based on factual responses provided earlier in the conversation, reported fewer issues.
However, its relevance score decreased compared to \leocw{p}{P}rompt \#7, which did not implement any filtering mechanism.

LLaVA-NeXT encountered significant difficulties with \leocw{p}{P}rompts \#7 and \#9, as well as with \leocw{p}{P}rompt \#6 from Experiment Two, suggesting it may struggle with multi-pass prompts.
This issue might stem from its smaller model size of 7B parameters instead of a larger 34B parameter model.
Our current hardware constraints prevent us from testing the larger model, presenting a challenge for future evaluations of open-source models.

\begin{table*}[hbt!]
  \caption[Experiment Three Results.]{Experiment Three Results:
  \leo{Despite having a high \fscore, in Prompt \#8, Copilot responded ``Misleading'' excessively on 99\% of the imbalanced 75 test cases ($N_{misleading}=63,N_{valid}=12$) while having a low relevance.}{}
  \leo{}{The \fscorespace (F1), precision (Prec.), and recall (Rec.) assess the LLMs' performance in an imbalanced setting ($N_{misleading}=63,N_{valid}=12$).}
  \leo{}{Relevance (Rel.) is the percentage of relevant responses.}
  \leo{}{An LLM's response is considered as relevant if the chart issue is reported as one of the pitfalls.}
  \leo{}{Factual Correctness (Fact.) measures the percentage of correct answers to the factual questions about the chart, \eg, chart type, title, axis, scale, and encoding.}
  \leo{Factual correctness}{It} does not apply to Prompt \#8 since it instructed LLMs to reply only to the summary question.}
  \squeezebetweentableandcaption
  \centering
  \includegraphics[width=\textwidth, clip, keepaspectratio, trim=0cm 26.3cm 0cm 0cm]{\getfigure{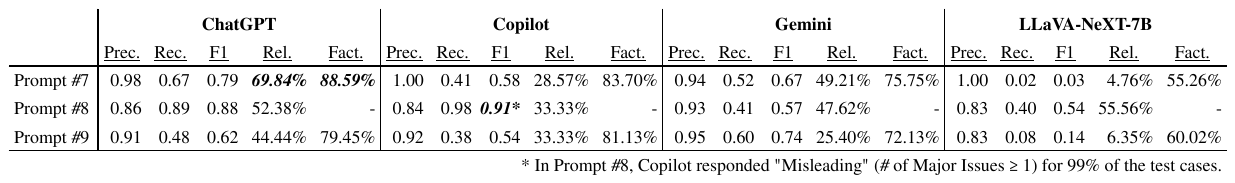}}
  \label{table:exp3}
  \squeezeaftertablethree
\end{table*}

\section{Discussion}

\textbf{LLMs demonstrated capabilities in chart comprehension.}
Given that the LLMs we assessed were not specifically trained to identify issues in charts, their performance exceeded our initial expectations.
This was particularly notable in their ability to analyze images (a task generally reserved for computer vision systems) and their comprehension of textual content within those images.
\leo{}{As shown in \cref{fig:factual}, LLMs accurately answered most questions regarding chart properties such as chart type, data type, data encoding, titles, and the use of 3D effects.}
\leo{}{However, they encountered difficulties with questions related to color counting and tick consistency checking, which is a known weakness in LLMs \cite{yang2023dawn}.}
\leo{}{Consequently, as illustrated in \cref{fig:issues}, LLMs performed less effectively on chart issues related to colors or inconsistencies in the taxonomy \cite{lo2022misinformed}, including overusing colors, indistinguishable colors, ineffective color schemes, and inconsistent tick intervals.}
\leo{It}{On the other hand, it} was surprising to see \leocw{}{that }these models \leocw{}{are }capable of differentiating between fictional and factual data, especially since such discernment is challenging for \leocw{standard}{rule-based} algorithmic detection systems.
This capability was demonstrated in the identification of ``not data'' charts, where LLMs had shown a robust understanding of charts and accurately reported the clues of fictional data in the images.

\leo{\textbf{Chain of thoughts is the most effective prompting strategy.}}{\textbf{Chain of Thought is the most effective prompting strategy in our experiments.}}
Over the course of \leocw{}{the }three experiments, we explored various prompting strategies to optimize LLM performance.
\leo{}{The intermediate output aided the model in building an understanding of complex misleading elements from simple chart properties.}
\leo{}{The concept of misrepresentation is challenging due to the logical dissonance of graphical integrity--the chart's geometry not reflecting the textual information on the chart.}

The CoT strategy emerged as the most effective, yet as we expanded the range of issues for detection, the prompts' length increased significantly.
To address this, we tested three alternative strategies: (1) dividing the prompt into a conversational format, (2) incorporating reasoning steps without explicitly including them in the output, and (3) dynamically generating prompts based on previous responses.
Our findings indicated that \leocw{p}{P}rompts \#4 and \#8, which did not include reasoning steps in the output, led to over-reporting issues.
Based on our experiments, \leocw{p}{P}rompt \#7 was identified as the most effective.
However, expanding the scope to encompass more issues poses a challenge, as it would significantly increase the prompt's length and complexity, potentially leading to lengthy response times and extensive outputs.

\leo{}{To address the scalability issue, alternative approaches should be considered.
Leveraging the LLMs' ability to generate \cite{do2023llms} or retrieve prompts \cite{rubin2022learning} from a prompt pool according to the input chart image and the conversational contexts could reduce prompt size while maintaining relevance.
The prompt pool could be built from an example dataset synthesized from previous work on taxonomy \cite{lo2022misinformed} and visualization guidelines derived from different visualization linters \cite{chen2021vizlinter, mcnutt2018linting}.}

\leo{}{Recent developments in prompt engineering include the agentic approach.
The mixture of agents (MoA) approach--utilizing multiple LLMs (with homogeneous or heterogeneous weights), each taking different roles in a conversation--has demonstrated improvement over the Chain of Thought approach \cite{wang2024mixture}.
Our experiments demonstrated consistent chart comprehension capabilities across different LLMs.
Applying the MoA and retrieval-based approaches may be promising next steps.}

\textbf{A benchmark dataset is urgently needed.}
Our evaluation efforts were limited by two main factors: (1) the absence of a benchmark dataset and (2) the need for refined evaluation metrics.
Although prior research has addressed the detection of misleading charts, there lacks a shared dataset and metrics for benchmarking purposes.
Consequently, we compiled our dataset of misleading and valid charts from existing studies on using visualizations on the internet.
This study serves as a foundation for subsequent research, facilitating comparison and overcoming the obstacles identified in our three experiments.

\section{Conclusion}

In this research, we explored the application of multimodal LLMs to the challenge of automatically detecting misleading charts.
Through a series of three experiments, which progressed from an initial set of five types of issues to an eventual examination of 21 issues, we identified effective strategies for prompting multimodal LLMs.
This progression allowed us to refine our approach and understand the nuances of utilizing LLMs for this specific task.
Each experiment contributed to a deeper understanding of how LLMs interpret and analyze charts and data, revealing both the capabilities and limitations of these models in the context of misleading visualizations.
Our findings unveil the potential and capabilities of LLMs in augmenting traditional methods for chart analysis and identifying misleading elements in charts.
The \leocw{drastic}{significant} development of LLMs provides a promising direction for further research in developing tools to support critical thinking on data interpretation and visualizations.

\bibliographystyle{abbrv-doi-hyperref}

\bibliography{template}

\end{document}